# Purely electronic nanometallic ReRAM

Yang Lu, Jung Ho Yoon, Yanhao Dong, and I.-Wei Chen

Yang Lu, University of Pennsylvania, USA; yanglu1@seas.upenn.edu

Jung Ho Yoon, University of Massachusetts Amherst, USA; yjh1309@umass.edu

Yanhao Dong, Massachusetts Institute of Technology, USA; dongyh@mit.edu

I.-Wei Chen, University of Pennsylvania, USA; iweichen@seas.upenn.edu

Resistance switching random access memory (ReRAM), with the ability to repeatedly modulate electrical resistance, has been highlighted as a feasible high-density memory with the potential to replace negative-AND (NAND) flash memory. Such resistance modulation usually involves ion migration and filament formation, which usually lead to relatively low device reliability and yield. Resistance switching can also come from an entirely electronic origin, as in nanometallic memory, by electron trapping and detrapping. Recent research has revealed additional merits of its mechanism, which entails smart, atomic-sized floating gates that can be easily engineered in amorphous Si, oxides, and nitrides. This article addresses the basic ideas of nanometallic ReRAM, which may also be a contender for analogue computing and non-von Neumann-type computation.

**Turning insulating thin films into nanoconductors and resistance switching random access memory**

A nanometallic memory is a purely electronic two-terminal device that offers fast switching speeds, small switching voltage, long retention time, and good durability. It is built on thin films of amorphous insulators and semiconductors, with a three-dimensional (3D) conducting network or channel embedded within and gated by one or more nearby trapped charge.[1–3] This trapped-charge memory is put into the high resistance state (HRS) when the trapped electrons block the channel via long-range Coulomb repulsion. Conversely, it enters the low resistance state (LRS) when the traps empty and the channel reopens. An as-fabricated nanometallic resistance switching random access memory (ReRAM) is in the LRS and nominally a conductor. But this only holds for thin films, typically less than ~10 nm thick.[2,3] Thicker films, like their bulk counterparts, tend to remain insulators and semiconductors, which do not switch unless first forming filaments by ion migration as in most other resistance switching random access memory (ReRAM). Thus, nanometallic ReRAM is distinctly a nanodevice.

Trapped-charge memory usually suffers from the voltage–time dilemma, which suggests that it is improbable for the memory to satisfy the three fundamental specifications for a memory—a small switching voltage, a fast switching time (speed), and long retention time. The argument is based on a simple theoretical consideration— an energy barrier that separates two memory states has only two independent characteristics—barrier height and width, so it cannot possibly satisfy three specifications. Often, a trapped electron occupies a higher energy state, so the escape barrier is lower than the capture barrier. However, nanometallic ReRAM is exempt from the dilemma because it has smart gates that autonomously reprogram the barrier to prevent detrapping.[2,3] As will become clear below, the reprogrammable barrier along with innate nanoconductivity and the size effect of this smart ReRAM are all rooted in the unique electron physics of disordered insulators.

**Negative-*U* centers**

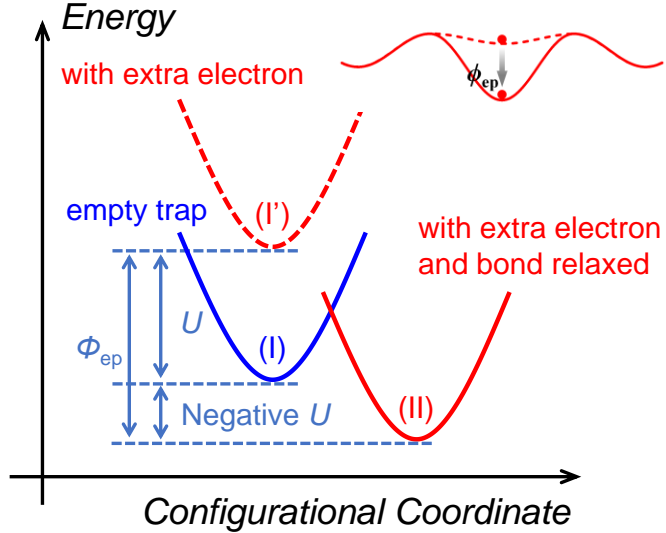

**Figure 1.** Energies of the ground state (blue), with an additional electron before (red, dashed) and after (red, solid) relaxation. Adapted with modification from Ref. 5. Inset at top right: Escape barrier for electron (dot) trapped around a negative-*U* center (solid curve) is higher than around a positive-*U* center (dashed curve) by $\phi_{ep}$.

Nanometallic memory uses negative-*U* centers, described below, to trap electrons.[3] As shown in the inset in **Figure 1**, after an electron is trapped at such a center, the energy of the trapped-electron state is lowered, so the escape barrier becomes higher, possibly even exceeding the capture barrier. Since there are now three independent characteristics, two barrier heights and one width, it is possible to satisfy three specifications. This explains why nanometallic ReRAM can resolve the voltage–time dilemma. The idea of negative-*U* centers was first proposed to explain the absence of unpaired electrons in amorphous chalcogenide semiconductors.[4] As illustrated in Figure 1,[5] starting with a reference state of *N*-electron system at I, the addition of another electron usually raises the system energy to I′ because of electron–electron repulsion. This energy increase, called the Hubbard *U*, is due to the on-site Coulomb repulsion between two electrons, but below we will extend the concept to the case of having one electron only when such electron must occupy a new orbital of a higher energy. In the single-site picture, *U* is positive and must be included in standard first-principles calculations (e.g., by the GGA + *U* method, GGA = generalized gradient approximation) of electronic structures if the material contains heavier atoms. However, if the $N + 1$-th electron induces the environment to relax to a new configuration II, then it inevitably lowers the energy by an amount $\phi_{ep}$, which reflects the crucial effect of the nucleus movement, hence "lattice" relaxation. Therefore, $\phi_{ep}$ caused by electron-induced bond relaxation is a measure of localized electron–phonon interaction. In chalcogenides, which have a soft matrix and large polarizability, such relaxation is so powerful that $\phi_{ep}$ exceeds $U$, making $U - \phi_{ep}$, or the effective $U$, negative and the final $N + 1$-electron state lower in energy than the initial $N$-electron state. These negative-$U$ centers are important for conductivity in chalcogenide memories.[6]

To facilitate lattice relaxation, the negative $U$ centers in insulators, which are not as polarizable as amorphous chalcogenides, must be locally soft to allow nucleus movement. It also helps if they can easily enter a symmetry-breaking configuration, which provides additional low-lying electron orbitals. This is illustrated in **Figure 2** for the one-electron case for three oxides using their GGA + $U$-computed electronic structures. In cubic $ZrO_2$ (Figure 2a), a reasonable band structure is obtained by assigning the Hubbard $U_{Zr} = 4$ eV to Zr, which shows fully occupied O-$2p$ levels and completely unoccupied Zr-$3d$ levels at ~3.3 eV higher, which is the effective $U$ at a lattice Zr site. At a defective Zr site (Figure 2b), surrounded by a Zr vacancy ($V_{Zr}$) and an O vacancy ($V_O$), the additional electron can occupy a new level at 2.0 eV; this is $U - \phi_{ep}$, so $\phi_{ep} = 1.3$ eV in this case. Moreover, keeping the same $V_{Zr}$–$V_O$ environment but placing the Zr at the saddle point (Figure 2b and Figure 2a left), which has lower symmetry and a softer environment, the additional electron now occupies a level only 0.3 eV above the O-$2p$ edge, which gives $\phi_{ep} = 3$ eV. Repeating the same calculation for cubic $CeO_2$ (Figure 2a center) with Hubbard $U_{Ce} = 5$ eV, we found a negative effective $U$ of $-0.3$ eV at the same saddle-point site despite a positive $U$ of 1.6 eV at the lattice Ce site. Finally, cubic $BaTiO_3$, which has the perovskite structure (Figure 2c and Figure 2a right) with $U_{Ti} = 4.36$ eV, has a positive $U$ of ~1.9 eV at the lattice Ti site, but a negative effective $U$ of $-0.3$ eV at the saddle-point site.

Therefore, even in relatively nonpolar oxide crystals, large $\phi_{ep}$ at soft, low-symmetry imperfections can yield slightly negative effective *U*.

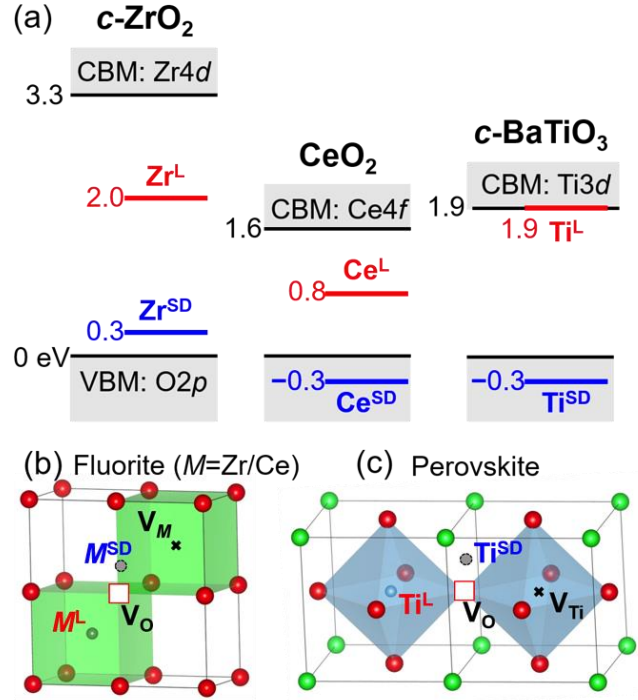

**Figure 2.** (a) Electron levels in (left) $ZrO_2$, (middle) $CeO_2$, and (right) $BaTiO_3$ according to GGA + Hubbard *U* calculations around a (b–c) cation next to a $V_O$–$V_M$ complex; such a cation at either lattice location ($M^L$) or saddle point location ($M^{SD}$). (b) $M$ = Zr or Ce in fluorite structure, or (c) $M$ = Ti in perovskite structure. (O indicated in red, Ba in green.) Effective *U* is defined as negative when electron energy lies below the valence band maximum (VBM). In a defect-free structure, the bandgap is between the VBM and the conduction band minimum (CBM). All numbers are in eV relative to VBM. Note: c-$ZrO_2$, cubic $ZrO_2$; c-$BaTiO_3$, cubic $BaTiO_3$; GGA, generalized gradient approximation; *U*, energy; $V_O$, oxygen vacancy; $V_M$, cation vacancy; $V_{Ti}$, titanium vacancy.

These findings naturally suggest that amorphous insulators are a fertile ground for finding negative-*U* centers. The local-structure landscape in disordered packing is diverse, and in some extreme terrains, there may be soft spots such as sites of large free volume. Under an external stress, some soft spots may deform irreversibly, which is why mechanical stress can "anneal" disordered packing, by "shaking it down." Such stressing removes $\phi_{ep}$ and consequently destabilizes the trapped electron, which may prompt electron escape. Although the magnitude of negative-*U* is difficult to calculate in disordered packing, since it requires sampling a large number of atomic configurations to faithfully and adequately survey the structural and electronic landscape, the idea has found relevance for nanometallic ReRAM, which utilizes amorphous insulators to take advantage of their negative-$U$ centers.

In the previously discussed example, we have taken a strict definition for negative-$U$, setting the reference energy level at the valence band maximum (VBM). In ReRAM, the insulator thin film is in contact with two electrodes, which sets the Fermi level halfway between their work functions and can be considerably higher than the VBM. To resolve the voltage–time dilemma, all that is needed is for state II in Figure 1 to lie below the Fermi level, because the Fermi energy is the energy of the electron before it enters the trap. Obviously, any substantially positive $\phi_{ep}$ will favor trapping when an electron is injected from an external source (e.g., an electrode) held at a higher energy, and any negative-$U$ center relative to the Fermi level will spontaneously accept a newly arriving electron that tunnels through from the source electrode. Some negative-$U$ centers may be so closely situated that the wave functions of their trapped electrons actually overlap. These "percolated" trapped electrons then form an extended state of conducting channel at an energy level well below the conduction band minimum (CBM). Therefore, negative-$U$ centers in a thin insulator film between the top and bottom electrodes can, in principle, provide not only trap sites for localized electrons, but also conducting channels of short lengths.

**Nanometallic transition**

To realize nanometallic ReRAM in which the trap sites serve as the floating gate that regulates the channel, one more key idea is needed—the localization length $\zeta$ of electrons.[1,2] By tuning the composition and thickness $\delta$ of the film in the device (**Figure 3**), one can obtain useful memory devices that satisfy $\zeta > \delta$ initially and $\zeta < \delta$ later after being switched to the HRS. The initial (LRS) $\zeta$ may be identified by the length of the channel; in the HRS state, the effective $\zeta$ is curtailed by the Coulomb blockade (due to electron-electron repulsion) imposed by the trapped electrons. Next, we illustrate this idea using two films with different bandgaps.

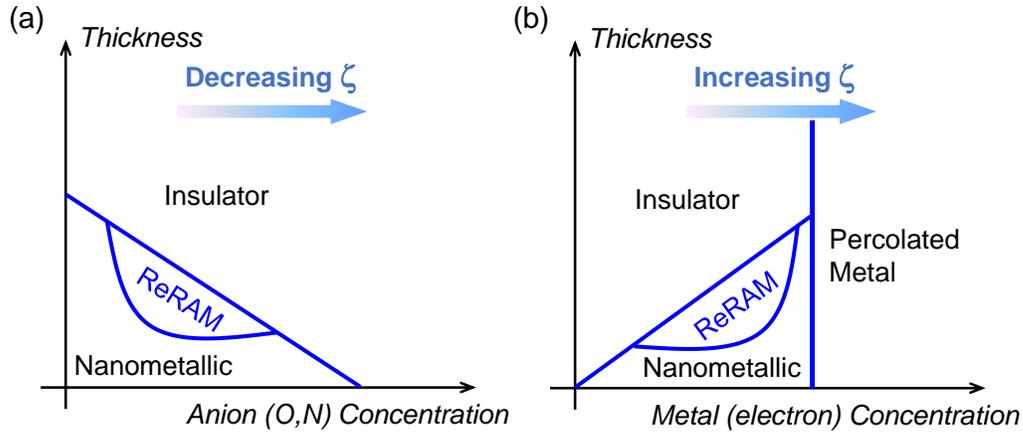

**Figure 3.** Generic composition–thickness maps. (a) Small bandgap insulator such as Si, (b) large bandgap insulator such as $Si_3N_4$. Note: $\zeta$, localization length of electrons.

*Amorphous silicon*

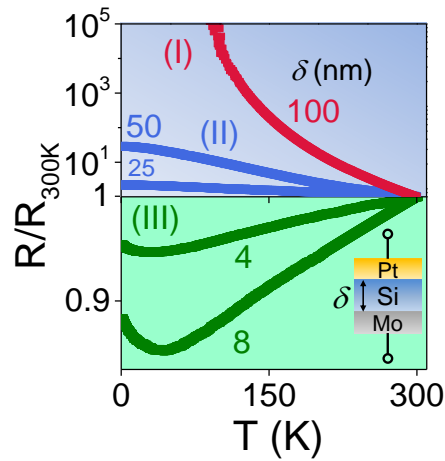

**Figure 4.** Nanometallic transition of resistance $R(T)$ with thickness with (I) insulating, (II) insulating plus low-temperature tunneling and (III) metallic characteristics, indicating a thickness-determined insulator (blue region) to metal (green region) transition. Note: $T$, temperature; $R_{300K}$, resistance at 300K; number next to curve indicates thickness.

Our first example of nanometallic transition is in amorphous silicon.[7,8] Amorphous silicon has a bandgap of about 1.8 eV, and even at the highest level of substitutional doping (by B and P), it cannot become a metal.[9] Therefore, bulk amorphous silicon always has zero conductivity at 0 K. However, as shown in **Figure 4**, [7,8] undoped amorphous Si film is conductive when its thickness is below ~20 nm. This thickness-mediated transition, seen in all the materials listed in **Table I**, has been termed nanometallic transition. At all temperatures and in every way studied, the thin, conducting films behave like a metal.

| Compositions |
|---|
| $Si_3N_4$, AlN:Al,Cr,Cu,Ta,Pt |
| $SiO_xN_y$:Pt |
| $SiO_2, Ta_2O_5, Y_2O_3,$ $HfO_2, MgO, Al_2O_3$:Pt |
| Si:O, Ge:O, n-Si:O, p-Si:O |
| Si:N, Ge:N, n-Si:N, p-Si:N |

**Table I.** Compositions of nanometallic memory. In their bulk form, compounds before the colon are amorphous insulators; in nanometallic thin films, they form matrices in which the element(s) after the colon are atomically dispersed.

The nanometallic transition occurs because percolation of electron-filled negative-$U$ states creates a conducting channel that bridges the two electrodes, but since long-distance percolation is impossible given the few negative-$U$ sites, only thin films are rendered conducting by this mechanism. To obtain a better picture of the conducting paths, we studied the temperature and magnetic-field (up to 18 T) dependence of conductivity at low temperatures (down to 18 mK) where inelastic

> **Quantum correction to conductivity (QCC) and Aharonov–Bohm oscillation**
> The electron energy in a metal stays constant at low temperatures because inelastic scattering is rare. Such electrons experience quantum interference, which gives rise to quantum correction to conductivity (QCC). In three-dimensional (3D) conduction, $\Delta\sigma$, the conductivity change relative to a reference state (e.g., the conductivity at 0 K) follows $d\Delta\sigma/dT \sim T^{-1/2}$. In one-dimensional (1D) conduction, it follows $d\Delta\sigma/dT \sim T^{-3/2}$. When a magnetic field $B$ is applied, the magnetoresistance is independent of orientation in 3D conduction, but not in 1D conduction. Furthermore, when the 1D conducting path self-crosses to tie a "wire loop," the magnetic field will generate the Aharonov–Bohm effect, altering the quantum phase of electrons traveling on the loop thus undulating the magnetoresistance. The undulation period reflects the magnetic flux required to reach a flux quanta, $\Phi_o = h/2e$, where h is the Planck constant and $e$ the charge of an electron, and unless the wire is infinitely thin, there is a path difference (hence destructive interference) between the inner- and outer-loop perimeters, which causes attenuation. The Aharonov–Bohm oscillation thus provides definitive evidence for single-filament conduction and informs the conducting path (the loop area and the wire cross section.)[8]

scattering is so infrequent that conducting electrons maintain the same energy for a long period of time. The quantum interference of these coherent electrons gives rise to the characteristic quantum correction to conductivity (QCC) that is sensitive to the dimensionality and topology of the conducting paths (see the QCC and Aharonov–Bohm oscillation sidebar).[10,11] Studying QCC, we found conduction in amorphous Si occurs along a three-dimensional (3D) network that occupies a tiny fraction of the film. But like a cotton ball, the network can still fill the gap between the electrodes, so it provides highly uniform conduction on a coarser scale.

Amorphous silicon, however, does not exhibit resistance switching whether undoped, *p*-type doped, or *n*-type doped. To make it into a ReRAM, anion (O or N) doping by either supplying $O_2$ or $N_2$ during reactive sputtering or co-sputtering Si with an oxide ($SiO_2$, $Al_2O_3$,) or nitride ($Si_3N_4$, AlN) is needed (Table I). Anion doping lowers the VBM, increases iconicity, and narrows the energy bandwidth of the channel. It shortens $\zeta$ (Figure 3a) and makes it easier for the trapped charge to disrupt the channel traffic.

To prove that the trap sites are negative-*U* centers endowed with a large $\phi_{ep}$, we performed two pressure-switching tests (see the Pressure tests sidebar) on devices that were initially voltage-switched to the HRS. After applying hydraulic pressure to uniformly compress the devices, we found that many had switched to the LRS. Since the

> **Static and dynamic pressure tests**
> Static pressure is generated in a vessel filled with a liquid, in which bare samples enclosed in a sealed evacuated elastomer bag are immersed. The pressure of the liquid medium is pumped to 2 MPa–400 MPa over a period of 1–2 min before release. Dynamic pressure is generated by an electron bunch, **Figure S1**a, which is a ~30-μm fast-moving "fireball" of 1–5 nC 22 GeV electrons, which reach almost the speed of light. In this energy range, the electron's collision cross section is small (the stopping power, which is proportional to the cross section, is ~10–20 MeV/cm in typical materials). Therefore, the device in the form of a parallel capacitor with dielectric "filling" is "transparent" to the bunch. What the dielectric does "sense" from a "distant" bunch—one that passes by in 0.1 ps—is a huge 0.1 ps magnetic impulse (peak field $B \sim 50$ T.) As in a Navy magnetic gun, the impulse generates a magnetic pressure that pushes against the electrodes and causes devices as far as 0.5 mm away to switch, from HRS to LRS (Figure S1b.)

pressure was uniform and the devices were not connected to any electrical meter during the pressure test, there could not be any stress- or electromigration of atoms or ions. The same test was repeated at the Stanford Linear Accelerator Center using an electron bunch (as well as a positron bunch), which contains about 1 billion electrons (or positrons) traveling at close to the speed of light yet remaining closely spaced within a size of 25 μm, which provides a negative pressure lasting 0.1 ps, and with the same outcome.[7,8] These pressure transitions are vivid manifestations of electron–phonon interactions. The pressure causes the environment of the negative-$U$ center to distort, which reverses $\phi_{ep}$ to make the environment inhospitable to the trapped electron, so the electron is "squeezed out." (In Figure 1, it corresponds to mechanically forcing the configuration II back to I'.) Indeed, the $10^{-13}$ s time in the dynamic pressure test is just long enough to allow an atom to move around the same "lattice" site once, but not long enough to wait for thermal activation to hop to another site. Therefore, the electron–phonon interaction witnessed here is one of athermally forced local-bond distortion, instead of thermally activated lattice vibration. Because the previously discussed mechanism only holds when there is already an electron trapped at the negative-$U$ site, pressure cannot trigger the LRS-to-HRS transition. Indeed, we did not observe this in any nanometallic ReRAM.

*Amorphous $Si_3N_4$*

Amorphous $Si_3N_4$ is a large bandgap (~5–6 eV) insulator regardless of film thickness. However, this changes with electron doping by co-sputtering $Si_3N_4$ and a metal (e.g., Pt or Cr), which deposits atomically dispersed metal atoms in the amorphous matrix. As the dopant concentration reaches a few atomic percent, thinner films become conducting, some even becoming switchable (Figure 3b).[3] Their metallic LRS follows the 3D scaling law in QCC and exhibits isotropic magnetoresistance. But the presence of Pt, a heavy element, imparts a spin–orbit interaction that exerts an opposite interference effect on QCC, which also manifests in our experiment.[8] Again, the pressure-triggered HRS-to-LRS transition confirms that $\phi_{ep}$ operates at the negative-$U$ centers.

To rule out the possibility of filamentary conduction in Si:O/N ReRAM and $Si_3N_4$:Pt ReRAM, we used a small mechanical force to break each device into two.[12] Comparing the areas $A$ of severed pieces and their resistance values $R$ with the initial values $A_0$ and $R_0$ of the intact device, we found they all have similar resistivity. This confirms uniform resistivity, hence uniform switching, in both the LRS and the HRS. In contrast, a filamentary ReRAM such as the one made of $HfO_2$ (more on this later) behaves differently. For LRS, only one-half inherited the same resistance as the intact piece, while the other half were orders of magnitude more resistive because it had no filament.[12]

**Nanometallic ReRAM**

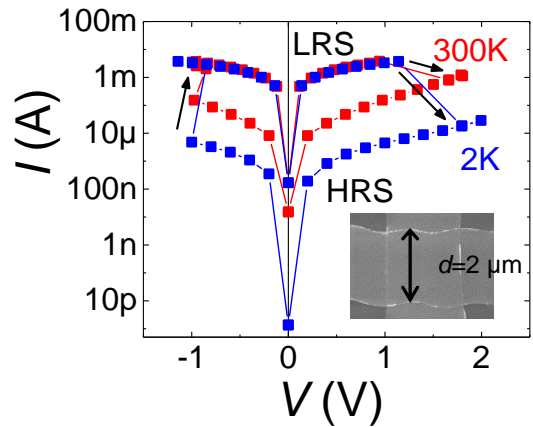

**Figure 5.** Current–voltage (*I–V*) curves at different temperatures showing non-volatile memory-like hysteresis and temperature-independent switching voltage. Inset: 2.5-µm crossbar device (scanning electron micrograph).[8] Note: LRS, low resistance state; HRS, high resistance state; arrows indicate the direction of the hysteresis loops; blue: 2K data; red: 300K data.

Generic bipolar current–voltage (*I—V*) curves[1–3,7] for ReRAM shown in **Figure 5** have the following features applicable to all the compositions listed in Table I. The LRS is typically Ohmic over the entire voltage range, which ends in either switching or breakdown. The HRS is initially Ohmic in the small voltage range, after which its resistance decreases rapidly (e.g., $R \sim V^{-4}$) and eventually approaches the LRS value. Remarkably, the switching voltage, at approximately ±1 V, is similar for ON and OFF switching, and is independent of the film thickness, temperature, and voltage-pulse width as long as it exceeds the resistor–capacitor (*RC*) charging time. For the 2.5-µm crossbar device shown in the inset in Figure 5, 1 ns switching time was demonstrated.[8] These features are consistent with a voltage-controlled transition between the metallic LRS state and the insulator HRS state and with the 0.1 ps dynamic-pressure-triggered HRS-to-LRS transition. At small voltage (≤±1 V), the HRS resistance follows the scaling law of variable-range hopping, increasing rapidly with a decreasing activation energy with decreasing temperature. The switching direction is distinctly dependent on the relative work function of the electrodes. ON switching to LRS occurs with electrons flowing to the low-work-function electrode, and OFF switching to HRS occurs with electrons injected from the low-work-function electrode. This is consistent with our picture of the electron energy states in nanometallic films, and it holds in approximately 25 tests pairing eight different electrode metals. (Ranked by work functions [in eV], they are Ta [4.25], Ag [4.26], Ti [4.33], Cr [4.5], Mo [4.6], Cu [4.65], TiN [4.65], $SrRuO_3$ [5], Au [5.1], Ni [5.15], and Pt [5.65].[13])

Most nanometallic memories investigated have uniform properties from cells to cells, as evident from the steep, nearly vertical slope in the Weibull plots of switching voltages and resistance values.[2,7] This is consistent with fracture tests, which verified electrical/microstructure uniformity down to the length scale of 10 μm.[12] Because the 3D network that spans between the electrodes is isotropic according to the magnetoresistance, area uniformity should extend over a length scale commensurate with the film thickness. Since we have demonstrated nanometallic ReRAM with a film as thin as 2 nm, spatial uniformity is achievable at the same length scale. Both the HRS and LRS are scalable (i.e., $R\sim1/A$) when switching is performed under the same current-density compliance, which is consistent with the picture of voltage-controlled switching.[14] These devices are highly stable. Devices stored for years maintaining their resistance memory are still switchable.[13]

**Switching filamentary memory electronically**

The nanometallic compositions in Table I are broad. Yet they all yield a similar ReRAM susceptible to a pressure-triggered HRS-to-LRS transition. We have already verified that the negative-*U* centers with ubiquitously large $\phi_{ep}$ in O/N-doped Si are caused by O/N doping, but are they also associated with metal dopants in metal-doped oxides and nitrides? To answer this question, we performed pressure tests on pristine atomic-layer-deposited $HfO_2$ and $Al_2O_3$, free of any metal dopant, with various electrodes.[8,15] These devices have a virgin resistance of 10–100 GΩ and are known to make excellent filamentary memories, becoming switchable after voltage forming, with their filamentary nature verified by fracture tests.[12] However, both the static and dynamic pressure tests found many of these HRS devices can pressure-switch to the LRS. Moreover, many virgin 10–100 GΩ devices also became conductive after the static and dynamic pressure tests, some with a pressure as low as 2 MPa.[8,15] These "pressure-formed/switched" LRS devices can be subsequently switched to the HRS (and going further through the switching cycles repeatedly) by a voltage, at both 300K and 2K, and their switching characteristics are indistinguishable from those of voltage-formed/switched devices. Therefore, there is no question that a large $\phi_{ep}$ and negative-$U$ centers do exist in undoped amorphous $HfO_2$ and $Al_2O_3$.

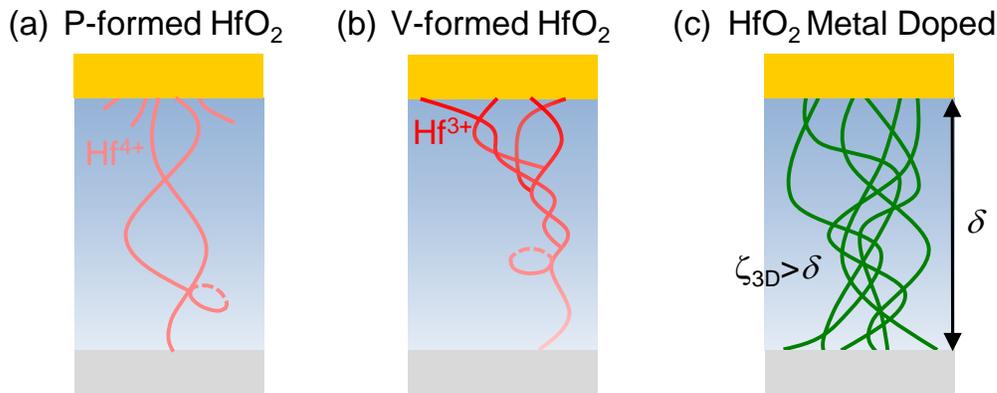

**Figure 6.** Conductive paths in three resistance switching random access memory (ReRAM): (a) $HfO_2$, pressure (P) formed, forming a self-crossing filament without reducing $Hf^{4+}$, (b) $HfO_2$, voltage (V) formed, forming a self-crossing filament, some incipient filaments and reduced $Hf^{3+}$. and (c) $HfO_2$, metal doped, with extensive 3D conducting network. Note: $\zeta_{3D}$, localization length in 3D electron conduction; δ, film thickness.

The LRS devices, whether voltage- or pressure-formed, are metallic at all temperatures. However, their QCC follows the 1D scaling law instead of the 3D one as followed by Si:O/N and $Si_3N_4$:Pt ReRAM. Remarkably, they also exhibit oscillatory magnetoresistance of the Aharonov–Bohm effect,[8] which is an unmistakable signature of quantum interference of coherent electrons in a single 1D conductive loop (**Figure 6**a and Sidebar: QCC and Aharonov–Bohm oscillation.) The period and the attenuation of the oscillation allow the determination of the loop radius, 7 nm, and the path cross section, 0.9 nm, in a pressure-formed $HfO_2$ ReRAM. For voltage-formed devices, we again witness the oscillations at higher field whereas the low-field magnetoresistance takes on a cusp shape, indicative of a spin–orbit interaction.[8] Therefore, while voltage forming has created the same 1D, self-crossing conducting paths (Figure 6b), it also comes with collateral damage in the form of reduced metal cations whose outer electrons give rise to the spin–orbit interaction.

The same pressure effect on these filamentary ReRAM implies their switching is at least partly controlled by the negative-$U$ centers and trapped charge. The fundamental difference between filamentary memory and nanometallic memory stems from their different bandgaps (Si: 1.8 eV, $Si_3N_4$: 5 eV, $HfO_2$: 5.7 eV), wider in the former, which is known to negatively correlate with the wave function overlap and the bandwidth of the channel. In filamentary ReRAM, because of the large bandgap and limited wave function overlap, the virgin film is typically far too thick to allow electron-filled negative-$U$ centers to percolate to form a long enough conductive filament. But such a filament does form occasionally after pressure forming, and quite frequently after aggressive voltage forming because of dielectric breakdown. In electron-doped compositions of Table I, the virgin film has a smaller bandgap and more electrons to allow many conductive pathways to build a 3D network as shown in Figure 6c, which is the case of nanometallic ReRAM. Silicon, because of its small bandgap, also forms a 3D network, but its bandwidth is too large for the trapped charge to disrupt conduction. Therefore, silicon requires bandgap tuning via anion doping to make it switchable.

**Summary**

Taking advantage of the extreme deformability and electron affinity at soft spots in a disordered packing, nanometallic ReRAM is an expressly engineered purely electronic ReRAM, with a 3D conductive network embedded in an insulator regulated by trapped electrons of autonomously programmed retention stability, thanks to electron-phonon interaction that lowers the energy of the trapped electron at the soft spots. Nanometallic ReRAM can be fabricated from many compositions in a very large material universe. These include silicon doped with oxygen or nitrogen that are completely process-compatible with integrated circuit technology, which gives nanometallic ReRAM a competitive advantage. The nanometallic nature is eventually suppressed by too wide a band gap, which is when nanometallic switching gives way to filamentary switching. Regardless type, simple methods can help clarify ReRAM characteristics—pressure tests to detect nonionic switching and fracture tests to confirm filamentary switching. Finally, methods are now available to remove the load effect to reveal the true voltage,[14,16] area,[14] temperature, and magnetic-field scaling[8,17] in these two-terminal devices.

**Acknowledgments**

I.W.C. would like to thank Y. Wang (Western Digital-Hitachi), A.B.K. Chen (Intel), (P.) X. Yang (Western Digital-Sandisk), S.-G. Kim (SK Hynix), J. Lee (SK Hynix), and B.J. Choi (Seoul University of Science and Technology) for contributions to developing nanometallic ReRAM. Thanks also to I. Karpov (Intel) for his interest and discussions during the course of the research, and NSF (DMR-1409114, and DMR-11–20901 for LRSM facilities), SRC (2012-IN-234) and FAME-STARnet (an SRC program sponsored by MARCO and DARPA) grants for current and past financial support.

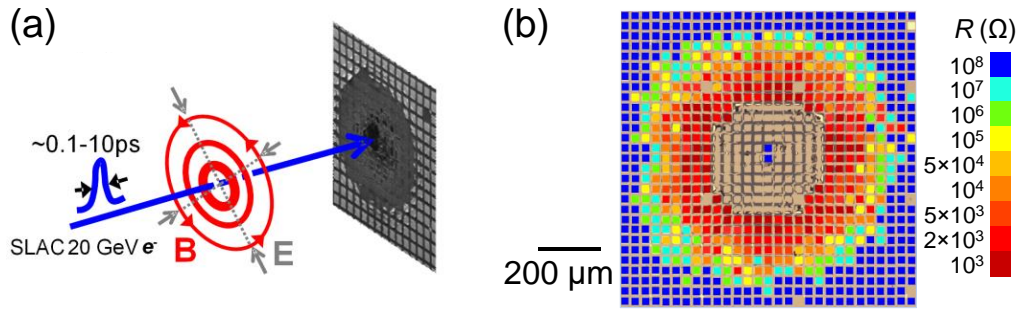

**Figure S1. Electron bunch delivers a magnetic pressure to switch ReRAM.** [3,8] (a) Schematic of electron bunch, with electrons bunched into a small size, which pass any given spot on the blue path in 0.1-10 ps, during which there is an intensive set of magnetic and electric field around the spot. The bunch hits the sample once, which results in (b) switching in previously set high resistance state (blue). Each "square" is one cell, about the same size as the electron bunch. Note: SLAC, Stanford Linear Accelerator Center; *B*, magnetic field; *E*, electric field; *R*, resistance.

**Yang Lu** received his BS degree in microelectronics in 2012 from Peking University, China, his MS degree in electrical engineering and his PhD degree in materials science and engineering, both in 2017 from the University of Pennsylvania. His current research includes nanometallic resistance memory and quantum electronic interference in nano-amorphous materials. Lu can be reached by email at yanglu1@seas.upenn.edu.

**Jung Ho Yoon** is a postdoctoral researcher at the University of Massachusetts Amherst. He received his PhD degree in materials science and engineering from Seoul National University, South Korea, in 2015. His research focuses on memristor materials and their novel applications. Yoon can be reached by email at yjh1309@umass.edu.

**Yanhao Dong** is a postdoctoral researcher at the Massachusetts Institute of Technology. He obtained his BS degree in materials science in 2012 from Tsinghua University, China, and his MS degree in materials science in 2014, his MS degree in applied mechanics in 2015, and his PhD degree in materials science in 2017, all from the University of Pennsylvania. He has eight papers published to date, covering atomic diffusion, ionic, and electronic defects; polarization; grain growth; and sintering. His current research focuses on functional oxides for energy applications. Dong can be reached by email at dongyh@mit.edu.

**I.-Wei Chen** is the Skirkanich Professor of Materials Innovation at the University of Pennsylvania (Penn), with former appointments at the University of Michigan and Massachusetts Institute of Technology (MIT). He obtained his BS degree in physics in 1972 from Tsinghua University, Taiwan, his MS degree in physics in 1975 from Penn, and his PhD degree in metallurgy in 1980 from MIT. He is an author of more than 100 papers in the *Journal of the American Ceramic Society* on oxide and nitride research, and his 200+ total journal papers have been cited more than 12,000 times representing an ISI *h*-index of 62. He holds 15 US patents and conducts current research on thin-film memory devices, energy ceramics, and drug delivery. Chen can be reached by email at iweichen@seas.upenn.edu.